\documentclass[graphics,floatfix, footinbib,tightenlines,nobibnotes, aps, prb, twocolumn]{revtex4-1}
\usepackage{amsmath,amssymb}
\usepackage{graphicx}
\usepackage{dcolumn}
\usepackage{bm}
\usepackage{braket}
\usepackage{subcaption}
\usepackage{verbatim}
\usepackage{float}
\usepackage{bbold}
\usepackage{color}
\usepackage{xcolor}
\usepackage{relsize}
\usepackage{amsthm}
\usepackage{enumerate}
\usepackage{soul,xcolor}
\usepackage[normalem]{ulem}
\usepackage[T1]{fontenc}
\usepackage[colorlinks=true ,urlcolor=blue,urlbordercolor={0 1 1}]{hyperref}

\newcommand{\beq}{\begin{eqnarray}}
\newcommand{\eeq}{\end{eqnarray}}

\newcommand{\bsp}{\begin{split}}
\newcommand{\esp}{\end{split}}

\newcommand{\be}{\begin{equation}}
\newcommand{\ee}{\end{equation}}

\begin{document}

\setstcolor{red}

\title{Quantum Hall spin liquids and their possible realization in moir\'e systems}
\author{
Ya-Hui Zhang}
\affiliation{Department of Physics, Harvard University, Cambridge, MA, USA
}
\author{T. Senthil}
\affiliation{Department of Physics, Masachusettts Institute of Technology, Cambridge, MA 01239, USA}
\date{\today}

\begin{abstract}
Recently Chern insulators with Chern number $C=1$ and $C=2$ in zero (or very small) magnetic field have been observed in two moire graphene systems: twisted bilayer graphene and  ABC trilayer graphene, both  aligned with a hexagonal Boron-Nitride (h-BN) substrate.   These Chern insulator states arise due to many body effects in the Chern bands of these systems when they are partially filled to a total  integer filling $\nu_T = 1,3$ (including spin and valley degrees of freedom).   A simple possible explanation is from Hartree-Fock mean field theory  which predicts  valley and spin polarization in the zero bandwidth limit, similar to the "quantum Hall ferromagnetism" in Landau levels.  Though valley polarization is implied by the existing experiments, the fate of the spin degree of freedom is not presently clear. In this paper we propose alternative valley polarized - but not spin polarized - candidates for the observed QAH effect.  For a valley polarized spinful Chern band at filling $\nu_T=1$, we describe  a class of exotic Chern insulator phases through spin-charge separation: charge is in a conventional Chern insulator phase with quantized Hall conductivity, while the spin forms disordered spin liquid phase with fractionalization, which we dub quantum Hall spin liquids. We construct a simple class of $Z_2$ quantum Hall spin liquid as analogs of  the familiar $Z_2$ spin liquid through slave fermion-schwinger boson parton theory.  Condensation of the spinon from the $Z_2$ quantum Hall spin liquid can lead to  a  quantum Hall antiferromagnet which is yet another, less exotic, candidate for the experimentally observed Chern insulator.  We offer several experimental proposals to probe the quantum Hall spin liquid  and the quantum Hall anti-ferromagnetic phases.  We briefly comment on the generalization to the filling $\nu_T=2$ and propose possible quantum valley Hall spin liquid without full spin polarization. Finally we also propose another class of QHSL using fermionic spinons, including phases supporting non-Abelian anyons.
\end{abstract}

\pacs{Valid PACS appear here}
\maketitle

\section{Introduction} 

 Recently moir\'e superlattices formed by Van der Waals heterostructures have been shown  to  be an excellent platform for strongly correlated physics. Observed phenomena  include correlated insulator\cite{cao2018correlated}, superconductivity\cite{cao2018unconventional,yankowitz2019tuning,lu2019superconductors} and (quantum) anomalous Hall effect\cite{Aaron2019Emergent,serlin2019intrinsic} in twisted bilayer graphene.  Spin-polarized correlated insulators\cite{Shen2019Observation, Liu2019Spin,Cao2019Electric}  and possibly superconductivity\cite{Shen2019Observation, Liu2019Spin} have been reported  in twisted bilayer-bilayer graphene. In addition, ABC trilayer graphene aligned with a hexagonal boron nitride (TLG-hBN) has been demonstrated to host gate tunable correlated insulators\cite{chen2018gate}, signatures of superconductivity\cite{Wang2019Signatures},  and a Chern insulator\cite{chen2019tunable}.

 Here we focus on the recent observation of Chern insulator phases in twisted bilayer graphene\cite{Aaron2019Emergent,serlin2019intrinsic} and in ABC trilayer graphene aligned with hBN\cite{chen2019tunable}. In these systems, there is one isolated Chern band per spin-valley flavor and the two valleys have opposite Chern numbers\cite{zhang2019nearly,chittari2019gate,zhang2019twisted,bultinck2019anomalous}. At odd integer filling $\nu_T=1$ or $\nu_T=3$, the valley is expected to be polarized because of the Coulomb exchange\cite{zhang2019nearly,bultinck2019anomalous}, resulting in a single spinful Chern band.  This is consistent with the experimentally observed hysteretic   anomalous Hall effect  in these systems.   To explain the Chern insulator behavior, the simplest option\cite{zhang2019nearly,bultinck2019anomalous} is to further postulate that the spin is fully polarized as well. The spin polarization is indeed favored in the flat band limit as verified by numerical simulation\cite{repellin2019ferromagnetism}. Full spin polarization is taken for granted in several theoretical works to explain the moir\'e Quantum Anomalous Hall  effects\cite{chatterjee2019symmetry,alavirad2019ferromagnetism,fengcheng2019}. 

 However, it is important to recognize that the current experimental data does not address the issue of spin polarization one way or the other.   In these moir\'e systems, the valley polarization is an ising order parameter, and hence there will be domain formation leading to hysteresis. To an excellent approximation the spin however is fully $SU(2)$ invariant. Thus the observed hysteretic transport suggests valley polarization (also known as orbital ferromagnetism) but does not directly give any evidence of spin polarization. A very recent experiment\cite{tschirhart2020imaging} directly imaged the magnetization in twisted bilayer graphene aligned with hBN  using a nano-SQUID probe. The measured magnetization is larger than that expected from spin polarization by a factor of $2-4$. This supports the picture that the magnetization comes predominantly from valley polarization.  Indeed these experiments could be consistent even with the complete absence of spin magnetism.  Given this situation, there is room to contemplate Chern insulator phases with more interesting spin physics than the usually assumed full spin polarization,  and this  is the focus of the current paper.

For simplicity, we assume valley polarization and consider a  model with  a single isolated spinful Chern band at filling $\nu_T=1$. We consider the possible phase diagram when the bandwidth $W$ varies relative to  the Coulomb interaction strength $U$. In the limit $\frac{W}{U}\rightarrow 0$, spin polarization is expected because  of the similarity to the Landau level and has indeed been verified by numerical simulation\cite{repellin2019ferromagnetism}. In the limit $\frac{W}{U}>>1$, the ground state is a conventional Fermi liquid. The key problem we want to study is the fate at intermediate $\frac{W}{U}$.  From Hartree Fock mean field theory, one expects that spin polarization gradually decreases to zero upon increasing the bandwidth, leading to a metal with spin imbalance in the intermediate region.  We will challenge this scenario in the following.

First let us review a related problem  to gain more intuition.  Consider spinful electrons in a topologically trivial band  at filling $\nu_T = 1$.  In the presence of interactions, a useful model of such a system is a lattice Hubbard model, which has been intensely studied for several decades. In many cases the ground state is an anti-ferromagnetic ordered Mott insulator at filling $\nu_T=1$ for small $W/U$, while it is a Fermi liquid metal for large $W/U$ (in the absence of special nesting effects).  How doe the system evolve between these two limits? In the conventional Hartree-Fock ("Slater") picture, the charge gap in the insulator is induced by the spin order. With increasing $W/U$ the spin order decreases and the system eventually goes through  an intermediate anti-ferromagnetic metal, before reaching the paramagnetic Fermi liquid. However,  for small $W/U$  the charge gap in the Mott insulator is induced by Hubbard $U$ and is independent of spin order.  Most importantly,  an intermediate Mott insulator with a quantum disordered spin state (quantum spin liquid) is possible. Indeed a spin liquid phase has been observed in numerical simulation of Hubbard model on the triangular lattice\cite{shirakawa2017ground,szasz2018observation}, and is further supported by experiments on some quasi-two dimensional triangular lattice organic salts. These results suggest that the naive Hartree-Fock theory is not correct in the case of a topologically trivial band.  Now let us return to the problem of the spinful Chern band. Because of the Wannier obstruction, charge can not be localized and the small $W/U$ regime is not described by a pure spin model.  In the small $W/U$ limit, the charge gap should still be determined by $U$ and disordering of spin does not necessarily close the charge gap. After all, the ferromagnetism is disordered at finite temperature because of $SU(2)$ rotation symmetry in 2D, but the QAH effect persists to finite temperature\cite{serlin2019intrinsic}.  We might then question whether Hartree-Fock theory correctly describes the evolution of the phase diagram with increasing $W/U$.  In a trivial band with non-zero Berry curvature, previous theory shows that the ferromagnetism in the $W=0$ limit can be suppressed by antiferromagnetic exchange at order $W/U$\cite{zhang2019bridging}. Similar destruction of the quantum Hall ferromagnetism by kinetic term  may also be possible at integer filling of Chern band.

\begin{figure}[ht]
\centering
\includegraphics[width=0.95\textwidth]{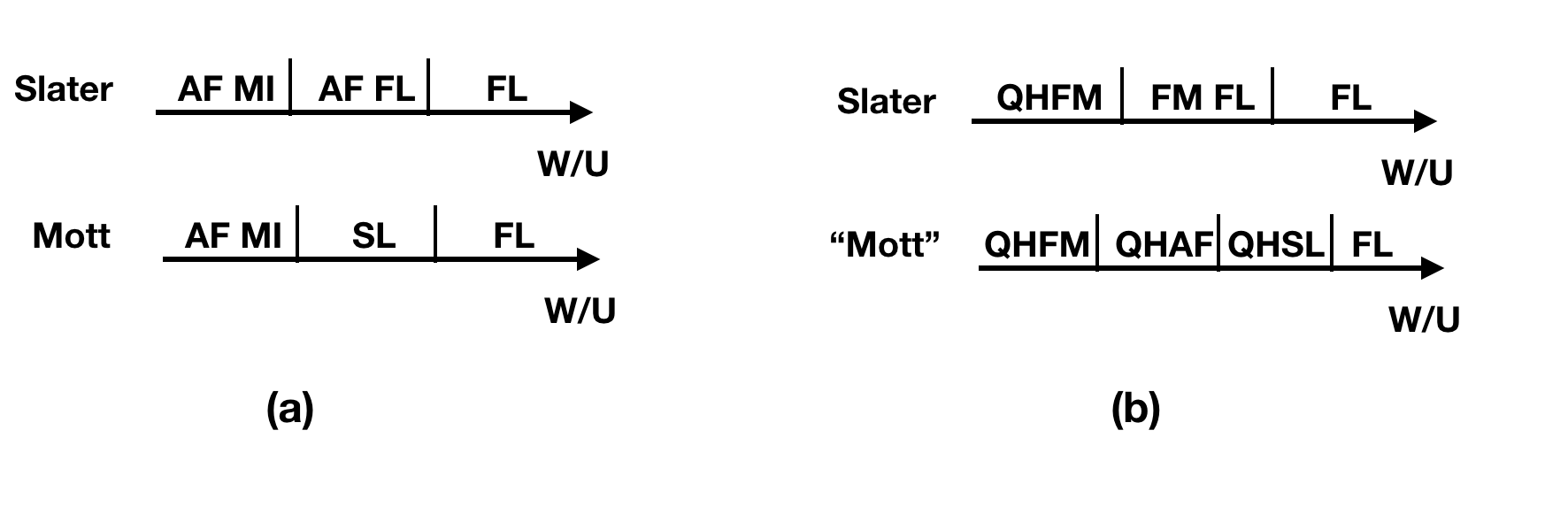}
\caption{Two types of phase diagrams for (a) spinful Hubbard model at $\nu_T=1$; (b) spinful Chern band at $\nu_T=1$. QHFM, QHAF and QHSL refer to quantum Hall ferromagnetism, quantum Hall antiferromagnetism and quantum Hall spin liquid respectively.  QHAF can be viewed as descendants of QHSL through a continuous transition. Within QHAF, we expect  that partial spin polarization coexist with antiferromagnetic order at momentum $\mathbf Q$.}
\label{fig:phase_diagram}
\end{figure}

In this paper  we will explore the possibility that  for intermediate $W/U$, the charge part remains as a Chern insulator with quantized Hall conductivity, but the spin is in a different state instead of a ferromagnet. This picture is illustrated in Fig.~\ref{fig:phase_diagram}. We consider both the possibility of antiferromagnetically ordered and of quantum spin liquid phases coexisting with 
quantized electrical Hall conductivity. We dub these states  quantum Hall antiferromagnets (QHAF) and quantum Hall spin liquids (QHSL).  Either of these states, as well as the simple spin polarized state, are consistent with current experimental observations. Thus it is important for future experiments to probe the spin physics and establish which of these states actually occurs. 

  From a conceptual point of view, one of  our main concerns in this paper will be to answer the  question: Which kind of symmetric correlated insulator with non-zero Hall conductivity can exist at integer filling of a spinful Chern band?   The analogous question of symmetric correlated insulator with zero Hall conductivity has been well studied in the context of Mott insulator of a topologically trivial band. In the Mott insulator, the charge degree of freedom is frozen at low energies and we can focus on just the spin degree of freedom to study a pure spin model obtained from a $t/U$ expansion.  This framework is not possible anymore in a Chern band, as the charge can not be frozen due to the Wannier obstruction. This is illustrated by the well studied quantum Hall ferromagnets in a Landau level of spinfull electrons: there due to a non-zero quantized Hall conductivity, topological spin textures (like skyrmions)  need to carry charge because they  are  felt by electrons as a magnetic flux\cite{girvin1999quantum}. As well see, quite generally, the spin and charge  degrees of freedom are  entangled in a non-trivial way in  a correlated insulator with a  Hall effect at integer filling.  Thus a QAH insulator without spin order is  different from the familiar spin liquid.  To the best of our knowledge, this class of phase has not been explored before and the various QHSL and Composite Fermi Liquid  phases constructed in the current paper  is a first description of some of the possibilities in this new direction.  As a secondary focus we will also describe some interesting properties of Quantum Hall Antiferromagnetic states which may also be relevant to  experiments.

Note that the ``quantum Hall spin liquid" really refers to a class of phases, which may be further divided to several categories (just like that there are many different `ordinary' spin liquid phases).  As explained above, by a QHSL, we mean a phase which has quantized integer Hall conductivity $\sigma^c_{xy}$ while its spin is in a disordered phase (very likely there exist fractionalized spinon excitation).  The notion of QHSL can be easily generalized to fractional quantum Hall spin liquid (FQHSL) which has fractional charge Hall conductivity. FQHSL should have both fractional charge and fractional spin. Spin rotation invariant FQHE states have been proposed before. The FQHSL proposed by us can be viewed as a new class of phases in this category. We will construct several simple QHSL and FQHSL phases explicitly, as well as the simpler QHAF phases. But we do not attempt to do a classification and more sophisticated QHSL phases beyond this paper definitely are possible.

Spin liquids have been discussed before in the context of moir\'e systems\cite{zhang2019bridging,wu2019ferromagnetism,zhang2020spin,kiese2020emergence}. However, these studies focus on the case with topologically trivial bands and the proposed spin liquids are the same as the ones studied in  traditional solid state systems.   In contrast, the QHSL phase discussed here should not be viewed as a subcategory of the familiar spin liquid. Instead, QHSL is a new class of phase, which shares properties of the both spin liquid and quantum Hall phases.  

Let us  highlight some attractive features of QHSL.  (I) Charge is not completely frozen in the QHSL, unlike the usual Mott insulator. In a gauge theory description, the internal flux is constrained to carry charge.  For example, in the $Z_2$ QHSL, the $m$ particle (the vison) carries charge $Q=\frac{C}{2}$ and fractional statistics. In the analogous state of "spinon Fermi surface" with $U(1)$ gauge field, the internal flux carries charge, making the phase compressible. (II) There is chiral electrically charged edge mode as implied by the quantized Hall conductivity.  Therefore the QHSL phase may be easier to detect than the traditional spin liquids, where all of excitations are neutral. (III) $Z_2$ QHSL realizes the same kind of quantum topological order present in  FQHE states but  at integer filling. The generalization to a xlosely related state - which we dub a quantum valley Hall spin liquid (QVHSL)-  realizes a fractional topological insulator at integer filling.  The correlated insulator at integer filling seems to be quite robust in moir\'e systems and the realization of fractional state at integer filling in QHSL may be easier than the usual proposal at fractional filling.  (IV)As the anyon in QHSL carries charge, doping a QHSL may create an anyon gas.  Especially in the case of QVHSL, we argue that an anyon superconductor can naturally emerge from doping. If this is true, the system is promising to search for parafermion mode\cite{alicea2016topological} by coupling a superconductor to the QVHSL insulator.

\onecolumngrid

\twocolumngrid

\section{General framework} 
\label{gf}
Consider a lattice model with spinful electrons $c_{i\sigma }$ (where $i$ is a lattice site and $\sigma$ is the spin component) such that there are 2 bands with equal and opposite Chern numbers $\pm C$. We assume that the lower band has positive Chern number and is well separated in energy from the other band. We consider a situation where the lattice filling $\nu_T = 1$ so that the lower band is half-full.  A concrete example (with $C = 1$) is provided by the spinful Haldane Hubbard model on the honeycomb lattice at this filling. 

We begin by using the standard slave fermion parton $c_{i\sigma}=f_i b_{i;\sigma}$, where $f$ is a slave fermion which carries the electrical charge while $b_{\sigma}$ is an electrically neutral bosonic spinon. As usual there is an emergent $U(1)$ gauge field arising from the constraint that the number of fermions $n_{fi} = f^\dagger_if_i$ must equal the number of spinons $n_{bi} = \sum_\sigma b^\dagger_{i\sigma} b_{i\sigma}$ at each site $i$:  $n_{fi}=n_{bi}$ . The total filling of the fermions is $n_f=1$. Therefore we consider an ansatz where the fermion $f$  inherits the Chern number of the  underlying electronic band and  is thus in a band Chern insulator phase. We may then contemplate a number of distinct phases depending on the fate of the spinons which we describe below. 

\subsection{Quantum Hall Ferromagnet}
The simplest possibility is that the bosons condense at a wave-vector $\vec q = 0$, {\em i.e} 
\begin{equation}
\langle b_{i\sigma} \rangle = \Phi_\sigma
\end{equation} 
where $\Phi$ is a spinor that is independent of $i$. Clearly this condensation gaps renders the gauge fluctuations innoccuous by the Anderson-Higgs mechanism. Further the order parameter 
$\Phi^\dagger \vec \sigma \Phi$ is gauge invariant, spatially uniform, and is a spin-triplet. Thus this state has ferromagnetic spin order. The charge response is determined by the $f$-fermions and is that of a Chern insulator. 

Thus we have just described the familiar Quantum Hall Ferromagnet which could equally well be understood within the usual Hartree-Fock theory. The purpose of the slave particle description is that it gives ready access also to the states described below some of which cannot be described within usual Hartree-Fock. 

\subsection{Quantum Hall Antiferromagnet} 
If the boson condenses at a non-zero wave vector $q$, then in general both spin rotation and translation invariance are broken and we have antiferromagnetic order. Specifically consider a condensate 
\begin{equation}
\langle b_{i\sigma} \rangle = e^{i\vec Q \cdot \vec x_i} \Phi_{+\sigma} + e^{-i\vec Q \cdot \vec x_i} \Phi_{-\sigma}
\end{equation}
This condensation also kills the gauge field fluctuations, and spontaneously breaks the global spin rotation symmetry. The gauge invariant spin triplet order parameters $\Phi_{+}^\dagger \vec \sigma \Phi_-$, $\Phi_-^\dagger \vec \sigma \Phi_+$ live at wavevectors $\pm 2\vec Q$ respectively. Thus the spin ordering is that of an antiferromagnet.  Depending on the details of $\Phi_{\pm}$, both uniaxial and spiral ordering patterns can be so described.  Specifically if $\Phi_+ = e^{i\theta}\Phi_-$ the ordering is uniaxial while if $\Phi_+^\dagger \Phi_- = 0$, the ordering is a spiral. The charge response is again due to the $f$-fermions, and is that of a Chern insulator. Thus we have a description of a Quantum Hall Antiferromagnet. 

 It is interesting to ask about topological defects of the antiferromagnet. For the spiral ordering pattern, the defects are point-like $Z_2$ vortices. Unlike conventional spiral ordered magnets, here these defects will carry non-zero electric charge 
due to the non-trivial quantum Hall charge response. Indeed these $Z_2$ vortices evolve into the vison excitations (the $m$-particle) when the quantum Hall antiferromagnet  undergoes a phase transition to the proximate $Z_2$ quantum Hall spin liquid described in the next subsection. Our analysis below of the charge of the quasiparticles of this quantum Hall spin liquid also determines the charge of the $Z_2$ vortices in the spiral ordered antiferromagnet. 

\subsection{Quantum Hall Spin Liquid}
We may contemplate the possibility that, instead of condensing, the bosons form a state which is a spin paramagnet. A simple such possibility is to let the Schwinger boson $b_\sigma$ form  a paired spin singlet superfluid phase described (at the mean field level) by the condensate 
\begin{equation}
\langle \epsilon_{\sigma \sigma'} b_{i\sigma} b_{j \sigma'} \rangle = P_{ij} 
\end{equation}
Here $\epsilon_{\sigma \sigma'}$ is antisymmetric with $\epsilon_{12} = 1$.  In a mean field description, this condensate leads to a term in the Hamiltonian of the form 
\begin{equation}
	H_b=\sum_{\langle ij \rangle}\Delta_{ij} (b_{i \uparrow} b_{j\downarrow}-b_{i\downarrow}b_{j\uparrow})+h.c.
\end{equation}

Clearly this condensation preserved spin rotation symmetry.  Further (analogous to the usual description of $Z_2$ quantum spin liquids in frustrated magnets) it breaks the $U(1)$ gauge structure to $Z_2$. Thus this state is a quantum spin liquid with topological order. The charge response however is still that of a Chern insulator, and we have the promised  ``Quantum Hall Spin Liquid" state.  

In the rest of the paper we describe the detailed topological order and properties of such Quantum Hall Spin Liquid states. We will see that despite the similarity in the construction, the Chern number of the $f$-band changes the topological order from that of the $Z_2$ spin liquid. Though our initial discussion will use the parton construction above, we will later also use an alternate parton construction in terms of fermionic spinons and bosonic holons. This will enable us to access a different class of Quantum Hall Spin Liquid states.

\section{$Z_2$ QHSL: Eight-fold way \label{section:abelian_Z2_QHSL}}

The low energy gauge theory for any of the states described by the bosonic spinon parton construction of the previous section takes the schematic form 
\begin{equation}
	L=L_f[f,a+A]+L_b[b,-a]
\end{equation}
where $a$ is the internal $U(1)$ gauge field. $b$ and $f$ have opposite gauge charges.  We have also included a coupling to a background (`probe') $U(1)$ gauge field $A$.  It couples to $f$ which carries the global electric charge but not directly to $b$. 

As we take  $f$ to be  in a Chern insulator with Chern number $C$, we have
\begin{equation}
\label{finCins}
	L_f[f,a+A]=-\sum_{I=1}^C \frac{1}{4\pi}\beta_I d \beta_I-\sum_{I=1}^C \frac{1}{2\pi}(A+a)d\beta_I
\end{equation}

Specializing to the QHSL state,  the pair condensation of the Schwinger boson is described by the following action (from boson particle-vortex duality)
\begin{equation}
	L_b[b,-a]=-\frac{2}{2\pi}a d \alpha
\end{equation}

Taken together, we get
\begin{equation}
	L=-\frac{1}{4\pi}\sum_{I=1}^C\beta_I d \beta_I-\sum_{I=1}\frac{1}{2\pi}(A+a)d\beta_I-\frac{2}{2\pi}a d \alpha
	\label{eq:first_step_chern_simons}
\end{equation}

We can simplify the theory by integrating $a$ (an alternative formulation can be found in the Appendix.~\ref{append:alternative_derivation}).   This enforces $2\alpha=-\sum_{I=1}^C \beta_I$. Substituting $\beta_C=-2\alpha-\sum_{I=1}^{C-1}\beta_I$, we get:

\begin{align}
	L&=-\frac{4}{4\pi}\alpha d \alpha -\frac{2}{4\pi}\sum_{I=1}^{C-1}\beta_I d \beta_I\notag\\
	&-\frac{1}{4\pi}\sum_{I \neq J}\beta_I d \beta_J -\frac{2}{2\pi}\alpha \sum_{I=1}^{C-1}d\beta_I+\frac{2}{2\pi}A d\alpha
	\label{eq:final_chern_simons}
\end{align}

In the basis of $(\alpha, \beta_1,...,\beta_{C-1})$, we have a a $K$ matrix with dimension $C$. 

 For $C=1$,  the theory is
 \begin{equation}
 	L=-\frac{4}{4\pi} \alpha d \alpha+\frac{2}{2\pi} A d \alpha
 \end{equation}
which is  a simple $U(1)_4$ theory with charge vector $q=2$.  In terms of topological order, this basically describes the $\nu=\frac{1}{4}$ Laughlin state of a Cooper pair formed out of the underlying electrons. However, this does not mean that there is pairing between electrons. Once there is a spin gap, single electron can be integrated out and the topological order can be viewed as purely bosonic.  This feature is also shared by gapped spin liquids in a traditional Mott insulator. 

 For $C = 2$ we can diagonalize the $2 \times 2$ K-matrix by the transformation  $K\rightarrow W K W^T$  with 
\begin{equation}
W=\left(
\begin{array}{cc}
1 &-1\\
0&1
\end{array}
\right)
\end{equation}

The transformed  $K$ matrix  is 
\begin{equation}
K=\left(
\begin{array}{cc}
2 &0\\
0&2
\end{array}
\right)
\label{eq:C_2_K}
\end{equation}
with charge vector $q^T=(2,0)$.

Basically the $C=2$ $Z_2$ QHSL contain two copies of $U(1)_2$ theory. One of them is charged and the other one is neutral.

The $K$ matrix for $C>2$ does not have a simple diagonal form.  Nevertheless  we can understand the general structure of the QHSL with general integer $C$.   First,  $|Det K|=4$ is true for any $C$ and it is always a bosonic Abelian topological order with four different anyons\footnote{By bosonic topological order we mean that the topological ordered sector can be chosen so that all local operators are bosonic}. 

From a general classification scheme\cite{wen2015theory}, there are only 9 different bosonic abelian topological orders with four anyons (up to staking of the invertible $E_8$ topological ordered state).  From explicit calculation of anyon self statistics and mutual statistics, we find $C=0, 1,2,3,...,8,...$ QHSL correspond to eight of them (the exception is the double semion) which has chiral central charge $c=0, 1,2,...7\ \  mod\  8$.  The QHSL with $C$ and $C+8n$ are equivalent in terms of anyon braiding and fusion.  Therefore $Z_2$ QHSL offers an interesting realization of eight different Abelian topological orders with four anyons. In terms of topological order they are equivalent to the 8 Abelian topological superconductors in  Kitaev's 16 fold way\cite{kitaev2006anyons}. 

\begin{table}[H]
\centering
\begin{tabular}{cccc}
anyon & $Q$ & $\theta$\\ \hline
$1$ & $0$   & $0$ \\ \hline
$e$ & $1$    & $\pi$ \\ \hline
$m$ & $\frac{C}{2}$  & $\frac{C}{4}\pi $ \\ \hline
$\epsilon$ &  $\frac{C}{2}+1$    & $\frac{C}{4}\pi$ \\ \hline
\end{tabular}
\caption{Anyons in the $Z_2$ quantum Hall spin liquid with Chern number $C$. We can always attach the single fermion to change $Q$, $S$ and $\theta$. We choose to fix $S=0$.  If we combine a local electron, $e$ particle can be viewed as a neutral bosonic spinon with $S=\frac{1}{2}$.}
\label{table:Z_2_anyons}
\end{table}

As there are four anyons, we can always label them by $1,e, m, \epsilon$.  We find that the fusion is different for odd $C$ and even $C$.  For odd $C$, we find a simple anyon $m$ with statistics $\theta=\frac{C}{4}\pi$ and charge $Q=\frac{C}{2}$.  Any other anyon is a composite of $m$.  We find $m^2$ has $Q=1$ and statistics $\theta=\pi$. After attaching the local fermion, $m^2$ is equivalent to a neutral bosonic spinon and thus we label it as $e$.  In another word, for odd $C$ we have $e=m^2$, $\epsilon=em=m^3$ and $m^4=1$.   In contrast, for even $C$ we need two simple anyons.  This is obvious for the $K$ matrix shown in Eq.~\ref{eq:C_2_K} for the $C=2$ case.  We label the charged one as $m$: this has self-statistics $\theta=\frac{C}{4}$ and $Q=\frac{C}{2}$.  We label the anyon in the neutral sector as $\epsilon$, which has  $\theta=\frac{C}{4}$ and  $Q=0$.  It is easy to show that $m^2=1$ and $\epsilon^2=1$. Meanwhile $e=m\epsilon$ has $\theta=\frac{C}{2}\pi$ and $Q=\frac{C}{2}$, which can be identified as the bosonic spinon up to attaching of the physical electron.  Therefore the fusion for even $C$ is the same as that of the  usual $Z_2$ gauge theory.  A list of anyons for the $Z_2$ QHSL can be found in Table.~\ref{table:Z_2_anyons}.

Even if QHSL is not the ground state, it may still be viewed as a parent state of the ordered phases described in Sec. \ref{gf}.  Specifically comndensation of the bosonic spinon $e$ leads to a magnetically ordered quantum Hall state.  A general argument - familiar from the theory of phase transitions out of $Z_2$ spin liquids\cite{chubukov1994universal,sachdev2002ground} - can be made that second order transitions will only occur for condensation at a non-zero wave vector. The resulting phase is thus a Quantum Hall Antiferromagnet.    Because of the spin-charge separation nature of our construction, the spin nature does not influence the charge gap and the quantized Hall conductivity.  In our construction, the momentum $\mathbf Q$ of QHAF can be arbitrary.  In contrast, one may be able to construct a Chern insulator with anti-ferromagnetic order within simple mean field theory.  In this Slater determinant picture, the $\mathbf Q$ of the AF needs to be fine tuned to a certain value to fully gap out the Fermi surfaces.  Therefore QHAF descending from QHSL is essentially different from a simple slater determinant state derived from an instability of the Fermi liquid. 

Within the QHAF phase the ordering wave vector will generically evolve continuously as the parameters of the microscopic Hamiltonian are changed. Thus a possible natural evolution is that there is a continuous phase transition from a QHSL to a QHAF which may then give way - with further change of parameters - to a Quantum Hall Ferromagnet.

\section{Generalization of $Z_2$ QHSL}

\subsection{Fractional quantum Hall spin liquid}
The notion of QHSL can be easily generalized to the case of fractional filling. . We  focus on a $C=1$ Chern band at filling $\nu_T=\frac{1}{3}$. As before we assume that the valley is polarized but do not make assumptions about the spin. Here we offer the simplest Fractional Quantum Hall Spin Liquid (FQHSL)  Now within the parton construction we have used  both the $f$-fermions and the $b$-spinons are at a filling of $1/3$. Taking the fermions to inherit the Chern number as before, they can form a fractional Chern insulator at filling $1/3$. Their low energy effective action is now replaced  from that in Eqn. \ref{finCins} by 
\begin{equation}
L_f[f, a+A] =  -\frac{3}{4\pi} \beta d\beta + \frac{1}{2\pi} (A + a) d\beta
\end{equation}

If we now consider a state where the spinons form a paired spin singlet superfluid , we will get an effective action 
 \begin{equation}
	L=-\frac{3}{4\pi} \beta d \beta+\frac{1}{2\pi}(A+a)d\beta-\frac{2}{2\pi}\alpha d a
	\label{eq:first_step_fqhs_chern_simons}
\end{equation}

We then integrate $a$ to find 

\begin{equation}
	L=-\frac{12}{4\pi}\alpha d \alpha+\frac{2}{2\pi}A d \alpha
\end{equation}

which is topologically the same as $\nu=\frac{1}{12}$ Laughlin state for Cooper pair. However we emphasize again that the route to forming this state does not involve attractive interactions. 

\subsection{Quantum valley Hall spin liquid at $\nu_T=2$}
Finally we briefly discuss the possibility of a Quantum Valley Hall spin liquid phase at $\nu_T = 2$. 
For the narrow Chern bands in  moir\'e systems with both spin and valley degrees of freedom at $\nu_T=2$, a correlated  quantum valley hall (QVH) insulator with full spin polarization is the likely ground state in the flat band limit\cite{zhang2019nearly,repellin2019ferromagnetism}. Indeed spin polarization has been observed in twisted double bilayer graphene (TDBG)\cite{Liu2019Spin,Shen2019Observation,Cao2019Electric}.  Similar to our discussion for $\nu_T=1$, when we increase the bandwidth, the full spin polarization can be destroyed first before the charge gap is killed.  Then  a quantum valley Hall spin liquid (QVHSL) is also possible.  In the slave fermion parton construction $c_{i;a \sigma}=f^a_{i;} b^a_{i;\sigma}$ with the constraint $n_{i;f^a}=n_{i;b^a}=n_{i;a}$. On average $\langle n_{i;+} \rangle+\langle n_{i;-} \rangle=2$.  In this construction there are two decoupled $U(1)$ gauge fields for the two valleys: $a^a_\mu$.   We can put $f_+$ and $f_-$ in Chern insulator with Chern number $C$ and $-C$. $b^a_\sigma$ can still be in paired superfluid.  The resulting state is the quantum valley Hall spin liquid (QVHSL).  Topologically it is equivalent to a two dimensional fractional topological insulator\cite{stern2016fractional}, but now it is realized at integer filling.

\subsection{Quantum Hall valley liquid}
 Finally we briefly mention another possibility with more detail in Appendix \ref{app:qhvl}  . 
QHSL is proposed for system with spinful Chern bands. Here we consider a different case:  we have two valleys with opposite Chern number and the total filling is $\nu_T=1$.  In the moir\'e context this corresponds to assuming full spin polarization, but not necessarily full valley polarization. It is possible to construct time reversal broken states with topological order which show an anomalous Hall effect but where the valley polarization can be continuously tuned. We denote these states ``Quantum Hall valley liquids'".

\section{Holon metal and  anyon superconductor upon doping a QHSL or QVHSL\label{sec:doping}}

In this section we briefly discuss possible phases from doping the $Z_2$ QHSL.  It will depend on what is the cheapest charged excitation in the QHSL phase. In a QHFM phase, the cheapest charged excitation can be either generated by $c^\dagger$ or the skyrmion defect of the FM order parameter.  In the slave boson parton theory, we disorder the slave boson condensation to melt the FM order. After that, the standard particle excitation is generated by $f^\dagger$ and the topological defect of FM order now becomes the $m$ anyon, which corresponds to meron of FM order and carries charge $Q=\frac{C}{2}$. Different phases can be obtained from doping depending on which of the above two excitations is cheaper.  We will discuss these two cases separately in the following.

\subsection{Holon metal}

The simplest case is that the cheapest charge excitation is generated by the slave fermion $f^\dagger$.  Then doping will just change $n_{f}=1-x$ and the schwinger boson $b_\sigma$ remains in a paired superfluid phase.  In the final phase, $f$ forms a single Fermi surface with area $A_{FS}=-x$ mod 1.  This is a metallic phase as $f$ carries the physical charge. But the Fermi surface couples to a $Z_2$ gauge field and the phase is different from a conventional Fermi liquid.  This is a holon metal\cite{kaul2008algebraic} in which both the single electron and the spin are gapped. This holon metal still breaks time reversal symmetry and has a non-zero Hall conductivity.    A similar holon metal can be obtained from doping a quantum valley Hall spin liquid and it is now time reversal invariant.

The charge transport of the holon metal is exactly the same as that of a ferromagnetic metal. Especially, the Landau fan degeneracy is reduced.   In the experiments, such a reduction of Landau fan degeneracy has been observed close to correlated insulator in several moir\'e materials. Usually it is attributed to a ferromagnetic order in the metal phase.   However, a holon metal is also consistent with these data.  The best way to distinguish the holon metal and a ferromagnetic metal is through the single electron gap or the spin gap.

\subsection{Anyon superconductor}

In a more nontrivial situation, the cheapest charged excitation is the $m$ anyon in the QHSL.  This anyon corresponds to the meron texture (half skyrmion) in the QHFM phase.  Once entering the QHSL phase, $m$ anyon costs only finite energy and can be the cheapest charged excitation.  In this case, we will have an anyon gas with finite density of $m$ anyons upon doping.  The problem of anyon gas has attracted lots of attention and it was predicted that an anyon superconductor phase may be favored to minimize the kinetic energy\cite{fetter1989random,chen1989anyon}. 

Superconductor may be even more likely if we dope the QVHSL phase, which has the same topological order as a fractional topological insulator.  For simplicity, let us consider the case with $C=1$. The QVHSL is described by a K matrix $K=\left(\begin{array}{cc}4 &0 \\
0 &-4 \end{array}\right) $ with charge vector $q=(2,2)^T$.  The are two sets of $m$ particles.  $m_+$ is generated by $l=(1,0)$ and $m_-$ is generated by $l=(0,1)$. Both carry charge $Q=\frac{e}{2}$ and have fractional statistics. Following the previous arguments in favor of anyon superconductor, $m_+$ and $m_-$ can pair and forms a boson with charge $Q=e$, which can then move coherently and condense.  The resulting phase is a superconductor.  The realization of a superconductor from doping a fractional topological insulator is interesting.  It has been proposed that   fractional topological insulator edges coupled to superconductors can be used to engineer parafermion states\cite{cheng2012superconducting,lindner2012fractionalizing,clarke2013exotic}.   These ingredients may all be available  within the same device in these moire graphene systems.

\section{Composite Fermi liquids: parent of another class of $Z_2$ QHSL\label{section:CFL}}
In previous sections we constructed  $Z_2$ QHSL as analogs of the familiar $Z_2$ spin liquids using the slave fermion-Cchwinger boson parton theory.  For conventional spin models, the same $Z_2$ spin liquid can also be accessed through the slave boson-Abrikosov fermion approach. In the fermionic spinon approach, one can get a $U(1)$ spin liquid with a spinon Fermi surface, which can be viewed as a parent state of the $Z_2$ spin liquid.  In this section we construct a variety of QHSL states as descendants of analogous parent states. For this purpose, we explore the slave boson-Abrikosov fermion parton theory.  For even $C$, we find that spin unpolarized composite fermi liquids (CFL) are the analog of the spinon Fermi surface of usual spin models. QHSL states -  distinct from those constructed in previous sections - can be obtained through pairing of the composite Fermi surface in CFL.  Unlike the usual spinon Fermi surface states, CFL is metallic.

We use the standard slave boson-Abrikosov fermion parton: $c_{i;\sigma}=b_i f_{i;\sigma}$. As before $b$ and $f_\sigma$ couple to  a dynamical  $U(1)$ gauge field\footnote{Strictly speaking a spin$_c$ connection.} $a$. We have the average density constraint $\langle n^b_{i} \rangle=1$ and $\sum_\sigma \langle n^f_{i;\sigma} \rangle=1$. We assume the mean field ansatz of $b$ inherits that Chern number of the original Chern band. Then we need to decide the fate of the slave boson at $n=1$ for a Chern band with Chern number $C$. Because we are searching for states with physical Hall conductivity, we should put $b$ in some kind of a quantum Hall state.  For even C, the simplest possibility is that $b$ is in a bosonic integer quantum Hall (BIQHE) phase\cite{lu2012theory,senthil2013integer} with $\sigma^b_{xy}=C$. We will come to the case with odd C later.

For even $C$, we put $b$ in the bIQHE state. The fermionic spinons $f_\sigma$ partially fill a band with no Chern number, and hence  it is natural to expect them to form  Fermi surfaces. The effective action of the corresponding phase is
\begin{equation}
	L=L_{bIQHE}[b,A+a]+L_{FS}[f,-a]
\end{equation}
where $A$ is a background  $U(1)$ gauge field\footnote{Actually a spin$_c$ connection.}. 
The effective induced action in the bIQHE state is simply\cite{lu2012theory,senthil2013integer}
\begin{equation}
\label{biqhe}
	L_{bIQHE}= \frac{C}{4\pi}(A+a)d(A+a)
\end{equation}
Note that the bIQHE state has no net thermal Hall effect (and hence no gravitational Chern-Simons term). 

and
\begin{equation}
\label{fs_a}
	L_{FS}=f^\dagger_\sigma(\partial_\tau -\mu+ia_0) f_\sigma-\frac{\hbar^2}{2m^*}f^\dagger_\sigma(-i\partial_i+a_i)^2f_\sigma
\end{equation}

\subsection{Property of CFL}

When there is a spinon Fermi surface,  the effective action of the resulting state is simply given by the sum of Eqns. \ref{fs_a} and \ref{biqhe}. This is of the same general form as the Halperin-Lee-Read (HLR)  action for the half-filled Landau level. The difference is that the level of the Chern-Simons term is $C$ instead of $1/2$. Despite this difference, the essential properties will  be very similar to the CFL in the half-filled Landau level.  In particular  it is metallic.  

A difference from the standard HLR theory is that in our  case the spin is not dead and we may contemplate composite fermi liquid states with  partial spin polarization. One interesting question is what the fate is when the spin is fully polarized. Unlike in the Landau level, when spin is fully polarized, the size of the spinon Fermi surface is the same as the BZ and fermionic spinon $f_\sigma$ should be gapped.  In this case the only term in the final action is that in Eqn. \ref{biqhe}.  We may now shift $a = \hat{a} =  a+A$ to get\footnote{Note that $\hat{a}$ is an ordinary $U(1)$ gauge field, and not a spin$_c$ connection.}. 
 \begin{equation}
	L=\frac{C}{4\pi} \hat{a} d \hat{a}
\end{equation}
This describes a non-trivial topological order with abelian anyons but it lives entirely in the electrically neutral sector, In particular there is 
no topological charge response (the Hall conductivity is zero).

In a  magnetic field, as usual the CFL can show quantum oscillations and form Jain sequences, as is shown in Appendix.~\ref{append:jain_sequence}.

\subsection{A different class of $Z_2$ QHSL from pairing of CFL}

It is easy to show that another class of $Z_2$ QHSL can be obtained from pairing of the spinon Fermi surfaces of the CFL. With pairing of $f_\sigma$,

\begin{equation}
	L_f=-\frac{2}{2\pi}a d \alpha
\end{equation}
where $\alpha$ is introduced through bosonic particle vortex duality.

Combining with Eqn. \ref{biqhe} we get 
\begin{equation}
	L=\frac{C}{4\pi} \hat{a}  d\hat{a} -\frac{2}{2\pi}\alpha d  \left( \hat{a} - A \right) 
\end{equation}

In terms of $(\alpha,\hat{a})$, we have a $K$ matrix:
\begin{equation}
	K=\left(\begin{array}{cc}
	0&2 \\
	2&-C
	\end{array}
	\right)
\end{equation}
and charge vector $q^T=(2,0)$.

For $C=2n$ with odd $n$, we can diagonalize the $K$ matrix by 

\begin{equation}
	W=\left(\begin{array}{cc}
	\frac{C+2}{4}&1 \\
	\frac{C-2}{4}&1
	\end{array}
	\right)
\end{equation}

The new $K$ matrix is 

\begin{equation}
	K=\left(\begin{array}{cc}
	2&0 \\
	0&-2
	\end{array}
	\right)
\end{equation}
and charge vector $q^T=(\frac{C+2}{2},\frac{C-2}{2})$.

For $C=4n$ with $n\in Z$, 

\begin{equation}
	W=\left(\begin{array}{cc}
	n&1 \\
	1&0
	\end{array}
	\right)
\end{equation}

transforms the $K$ matrix to

\begin{equation}
	K=\left(\begin{array}{cc}
	0&2 \\
	2&0
	\end{array}
	\right)
\end{equation}
and charge vector $q^T=(\frac{C}{2},2)$.

Obviously $C=4n+2$ with $n\in Z$ corresponds to double semion while  $C=4n$ with $n\in Z$ corresponds to toric code.  Within each class, the charge vectors are different for different $C$, leading to different Hall conductivity.  These states have helical edge modes protected by charge conservation.

\section{Non-Abelian QHSL at odd $C$: another eight-fold way}

We turn to the case with odd $C$. In this case the slave boson needs to be in an exotic state to have Hall conductivity $\sigma^b_{xy}=C$. Taking $C=1$ as an example, it has been shown that a boson at $\nu=1$ is in a Pfaffian state.  Motivated this, we consider the CFL phase with slave boson in a Pfaffian state while the Abrikosov fermion $f_\sigma$ form spinon Fermi surfaces.

The low energy action is:
\begin{equation}
 	L=f^\dagger_\sigma(\partial_\tau -\mu+ia_0) f_\sigma-\frac{\hbar^2}{2m^*}f^\dagger_\sigma(-i\partial_i+a_i)^2f_\sigma+L_{Pf}[b,A+a]
 \end{equation}

A (generalized) bosonic Pfaffian with $\sigma^b_{xy}=C$ can be understood as (Abelian-TQFT$\times$ Ising)/$Z_2$, where Abelian-TQFT is an Abelian topological order described by a $K$ matrix. For $C=1$ it is just $U(1)_4$. For a generic odd $C$, we can construct it from parton theory $b=\psi_1 \psi_2$ and let $\psi_1$ be in an IQHE state with $\sigma^{\psi_1}_{xy}=C$ and $\psi_2$ form a $p+ip$ superconductor. The Abelian TQFT part is

\begin{align}
	&L^b_{Abelian-TQFT}[b,A+a]=-\frac{4}{4\pi} \alpha d \alpha-\frac{2}{2\pi} \alpha d \sum_{I=1,...,C-1} d \beta_I\notag\\
	&-\frac{2}{4\pi} \sum_I \beta_I d \beta_I-\frac{1}{4\pi} \sum_{I \neq J} \beta_I d \beta_J+\frac{2}{2\pi}(A+a) d \alpha
	\label{eq:b_abelian-TQFT}
\end{align}

This should be glued to the non-abelian Ising anyon theory with 3 quasiparticles $(1, \sigma, \psi)$ where $\psi$ is a fermion, and $\sigma$ is the Ising anyon. The gluing condition is that 
the $\sigma$ is bound to anyons in the abelian theory with odd $q_\alpha$, while $1$ and $\psi$ are bound to abelian anyons with even $q_\alpha$.

Similar to the discussions for even $C$, we consider a paired state  of the spinon $f_\sigma$, which leads to
\begin{equation}
	L_f=-\frac{2}{2\pi}a d \tilde \alpha
	\label{eq:f_pairing}
\end{equation}

The final state is a non-Abelian QHSL, which can be  decomposed as (Abelian-TQFT$\times$ Ising)/$Z_2$.  The Abelian-TQFT part is a sum of Eq.~\ref{eq:b_abelian-TQFT} and Eq.~\ref{eq:f_pairing}.

We make the redefinition: $\alpha=\alpha_1$, $\tilde \alpha =\alpha_1- \alpha_2$, the final result is 

\begin{align}
	&L^c_{Abelian-TQFT}=-\frac{4}{4\pi} \alpha_1 d \alpha_1-\frac{2}{2\pi} \alpha_1 d \sum_{I=1,...,C-1} d \beta_I\notag\\
	&-\frac{2}{4\pi} \sum_I \beta_I d \beta_I-\frac{1}{4\pi} \sum_{I \neq J} \beta_I d \beta_J+\frac{2}{2\pi}A d \alpha_1+\frac{2}{2\pi}a d \alpha_2
	\label{eq:QHSL_abelian-TQFT}
\end{align}

We find that this Abelian TQFT is decomposed to TQFT$_1$ $\times$ Toric Code. Here Toric code part is formed by $(a,\tilde \alpha_2)$.  Interestingly, TQFT$_1$ is the same as that in Eq.~\ref{eq:final_chern_simons} for the $Z_2$ QHSL constructed from slave fermion-Schwinger boson approach.  From $q_\alpha \alpha+q_{\tilde \alpha}\tilde \alpha=q_\alpha \alpha_1+q_{\tilde \alpha}(\alpha_1-\alpha_2)$,  we can get the transformation rulel for the charge of $\alpha_1,\alpha_2$: $q_1=q_\alpha+q_{\tilde \alpha}$ and $q_2=-q_{\tilde \alpha}$.   Then $q_{\alpha}=q_1+q_2$.  We know that an Ising anyon is bound to $q_{\alpha}=1$ mod 2. In terms of gauge fields in Eq.~\ref{eq:QHSL_abelian-TQFT}, the Ising anyon is bounded to $q_1+q_2=1$ mod 2.  The final TQFT of this non-abelian QHSL is (TQFT$_1$ $\times$ Toric Code $\times$ Ising)$/Z_2$ with the rule that the $\sigma$ particle of the Ising topological order is bounded to $q_1+q_2=1$ mod 2. 

The chiral central charge is $c=C+\frac{1}{2}$.  We can also choose the Ising part as anti-Ising, leading to (TQFT$_1$ $\times$ Toric Code $\times$ anti-Ising)$/Z_2$ with chiral central charge $c=C-\frac{1}{2}$.  For $C=1,3,5,7$, we can generate another 8 different QHSL with chiral central charge $c=\frac{1}{2}, \frac{3}{2}, \frac{5}{2}, ..., \frac{15}{2}$.  They are similar to the 8 non-abelian superconductor in Kitaev's sixteen fold way\cite{kitaev2006anyons}. However, the topological order here has more anyons because of the Toric code part (see Appendix.~\ref{append:counting_anyons}).

\section{Experimental Signatures}

 We now discuss some possible experimental signatures of QHSL or QHAF. To date, much of the information on moire graphene materials has come from transport experiments. But in near future measurement of optical conductivity should also be feasible.  In the following we will list some probes to distinguish a QAH insulator with and without spin-valley polarization.

A useful probe of spin polarization or lack thereof is to study the charge gap in the presence of an in-plane magnetic field $B_x$. Such a field certainly couples to the spin degree of freedom. It may also couple to the orbital degree of freedom but this depends on details\cite{lee2019theory}. Assuming the in-plane field predominantly couples to spin, the charge gap $\Delta$ measured in a transport experiment will depend on it, and will be a measure of the spin of the cheapest gapped charged excitation. Under Zeeman field, charge gap changes as $\Delta(B)-\Delta(0)=g_{eff}\mu_B B$. Next we discuss our expectation of $g_{eff}$ for various phases. For a phase with spin-valley polarized to be $-,\downarrow$, the particle-hole excitations can be grouped in the following three categories:  (I) spin flip $c^\dagger \sigma^+ c$;  (II) valley flip $c^\dagger \tau^+ c$;  (III) spin and valley flip $c^\dagger \tau^+ \sigma^+ c$.  Here we denote $\tau_a$ as Pauli matrix in valley space and $\sigma_a$ in spin space.  We consider a particle-hole excitation separated by a long distance and label the energy gap corresponding to the above three categories as $\Delta_1,\Delta_2,\Delta_3$.  At zero magnetic field, we expect $\Delta_2 \approx \Delta_3$ because of an approximate $SU(2)_+ \times SU(2)_-$ symmetry\cite{zhang2019nearly}.  Under a Zeeman field $- g \mu_B B S_z$\footnote{We choose the Zeeman field to couple to $S_z$ to simplify the notation. In experiment Zeeman field is generated by in-plane magnetic field $B_z$, as $B_z$ also couples to the valley},  these three charge gaps change differently because they carry different spin:  $\Delta_1(B)-\Delta_1(0)=g \mu_B B$, $\Delta_2(B)-\Delta_2(0)=0$ and $\Delta_3(B)-\Delta_3(0)=g \mu_B B$ with $g=2$.   The charge gap will be set by the smallest one among $\Delta_1, \Delta_2, \Delta_3$\footnote{Here we ignore the possibility that the cheapest charged excitation is a skyrmion, which has been found to cost larger energy\cite{fengcheng2019}}.  $\Delta_2$ should be slightly smaller than $\Delta_3$ because of a small inter-valley Hund's coupling\cite{zhang2019nearly}, which however may compete with a phonon-mediated interaction of the opposite sign\cite{bultinck2019anomalous}   but the comparison between $\Delta_1$ and $\Delta_2$ depends on microscopic details and is not clear. Therefore both $g_{eff}=2$ and $g_{eff}=0$ are possible for spin-valley polarized phase.  Next we turn attention to  a QHSL phase with $S=0$ ground state. The cheapest charged excitation can carry $S=\frac{1}{2}$ (for example, generated by the electron operator), or carry $S=0$ (for example, the $m$ anyon). In the former case, the  energy cost for excitation with $S_z=\frac{1}{2}$  changes as $\Delta(B)-\Delta(0)=-g \mu_B B S_z=-\mu_B B$ under the Zeeman field. As the ground state energy does not change under Zeeman field because it is in a singlet state,  the charge gap changes as $\Delta(B)-\Delta(0)=- \mu_B B$ and we have $g_{eff}=-1$. In the later case with spinless charged excitation, we expect $g_{eff}=0$.  In summary, $g_{eff}=2$ is a strong indication of spin polarized state. $g_{eff}=-1$ is evidence for a ground state with zero spin polarization. However, $g_{eff}=0$ seems to be consistent with both spin polarized and spin unpolarized phases.  For QHAF phase with partial spin polarization, $0<g_{eff}<2$ is also possible.

 A QAH insulator without full spin-valley polarization will also be reflected in optical spectrum. In experiments it is possible to measure the optical conductivity $\sigma(\omega)$ for the correlated insulator.  Usually $\sigma(\omega)$ will has a threshold determined by the charge gap $\Delta$. However, $\sigma(\omega)$ remains as zero for $\omega>\Delta$ for a spin-valley polarized state. This is because only excitation with zero spin and almost zero momentum can be excited in optics.  For a spin-valley polarized phase, the particle-hole  excitation needs to flip either spin or valley, which is dark in optics.  Therefore optical conductivity is suppressed for a spin-valley polarized phase even at energy larger than the charge gap.  In contrast, for a state without full spin-valley polarization, there can be charged excitation which carries zero spin and zero valley quantum numbers , and so it can be excited optically. Thus $\sigma(\omega)$ onsets above a charge gap.

 A more definitive experiment that can distinguish the QHSL from other quantum Hall states is to study edge tunneling. For the QHSL, single electron is gapped even at edge. Therefore there is no single electron tunneling at the edge. The leading process is the tunneling of a cooper pair.   In contrast, single electron gap is zero for both QHFM and QHAF.  Finally, for the QHSL with odd $C$, there is an excitation with half charge, which may be detected by shot noise.
 
Let us now comment on experimental situations which may favor QHSL states.  As emphasized in previous papers, in the flat band limit the QHFM is likely the ground state\cite{repellin2019ferromagnetism}. Thus to stabilize other states we need to imagine tuning the band width. This can be done  by tuning a perpendicular displacement field which (at least in some moire graphene systems) tunes the band width. Direct calculations of band width over displacement field in ABC trilayer graphene aligned with hBN and in twisted monolayer-bilayer graphene can be found in Ref.~\onlinecite{zhang2019nearly,zhang2019bridging,chen2020electrically}. Indeed, metal insulator transition tuned by displacement field has been observed in these systems\cite{chen2019tunable,chen2020electrically}. QAH effect is seen in a finite region of displacement field $D$ (for example, please see Fig.4 of Ref.~\onlinecite{chen2020electrically}). Close to the boundary of the QAH effect, the system is also close to a metal-insulator transition and the quantum Hall ferromagnetism framework is likely not valid anymore.  As we argue in the main text, spin polarization can be lost first before the charge gap closes when increasing $D$. Hence we expect a possible phase transition hidden inside the correlated insulator in the existing experiments.  It will be interesting to study the quantum anomalous Hall state closely as a function of both displacement field and an in-plane magnetic field to detect any possible spin phase transitions. An alternate route is to change the twist angle between the two graphene layers slightly away from the magic angle in the twisted bilayer graphene aligned with hBN\cite{serlin2019intrinsic}. In the range of twist angles where the QAH effect persists, it is possible that there is a region where we have either the QHAF or the QHSL state. 

The possibility of FQHSL or FQHAF states must be kept in mind for future experiments on TBLG/h-BN away from $3/4$ filling. In the devices studied in Ref. \onlinecite{serlin2019intrinsic} evidence of valley polarization was seen in a range of fillings that included $\nu_T = \frac{5}{6}$. At this filling (which corresponds to a filling of $2/3$ by spinful holes in a $C = 1$ band),  theoretical calculations  of Ref.\onlinecite{repellin2019chern} showed that a Fractional Chern Insulator state is possible. There is however a close competition between spin singlet and spin polarized states even in the flat band limit.  Away from the flat band limit, should a Fractional Quantum Anomalous Hall Effect be present, we expect that FQHFM, FQHAF, and FQHSL states will all be candidates. 

 Lastly we want to point out another promising platform to study QHSL physics, which has not been realized in the experiments. The idea is to study a bilayer system with two layers coupled together by the Coulomb interaction.  The layer degree of freedom can mimic a spin $1/2$ and the advantage is that the pseudospin conductivity can be measured through counter-flow transport.  A QHFM phase with the layer pseudospin polarized in the $xy$ direction has been verified in this way by experiment in quantum Hall bilayers\cite{eisenstein2004bose,liu2017quantum}.  One can try to add dispersion to the Landau levels by inducing moir\'e superlattice potential in this system by aligning hBN to the double graphene layers in the system of Ref.~\onlinecite{liu2017quantum}. Or alternatively, we can consider a double moir\'e layers as proposed in Ref.~\onlinecite{zhang2020electrical}. For trilayer graphene-hBN-trilayer graphene system, one can tune both graphene layers to have topological moir\'e band\cite{zhang2019nearly}. In this case there are eight flavors formed by spin-valley-layer. At $\nu_T=1$, a QAH insulator with spin-valley-layer polarization is expected. When tune displacement field, bandwidth increases and the layer polarization may be disordered as we proposed in the paper.  It is easy to distinguish QHSL and QHFM/QHAF phases in this setting up because the pseudospin conductivity corresponding to the layer can easily measured through counter-flow transport. For QHFM and QHAF phase, the pseudospin resistivity should be  zero as the easy-plane anisotropy favors the polarization in the XY direction.  However, if the layer pseudospin is in a similar phase as the $Z_2$ QHSL, the pseudospin resistivity is infinite because of the pseudospin gap.

\section{Conclusion}
In summary, we propose a new classes of topological phase which we call quantum hall spin liquid, which is a combination of quantum hall state and spin liquid state.  QHSL can be viewed as parent state for quantum Hall ferromagnetism and quantum Hall antiferromagnetism.  Recently quantum anomalous Hall effect has been observed in moir\'e systems. In addition to QHFM, QHSL and QHAF are also consistent with the existing experimental results. We suggest several future experiments to further distinguish them.

\textit{Note added} During the finalization of this paper, we became aware of other works\cite{kwan2020excitonic,stefanidis2020excitonic} on the QAH effect from a different perspective.

\section{Acknowledgment}
TS was supported by NSF grant DMR-1911666,
and partially through a Simons Investigator Award from
the Simons Foundation to Senthil Todadri. YHZ was supported by a Simons Investigator Grant (PI:Ashvin Vishwanath).  YHZ also thanks  ``Simons Collaboration on Ultra Quantum Matter Workshop'',  which was supported by a grant from the Simons Foundation (651440).

\bibliographystyle{apsrev4-1}
\bibliography{qhsl}

\onecolumngrid
\appendix

\section{Alternative derivation of the $Z_2$ QHSL: integrating $\alpha_I$ \label{append:alternative_derivation}}
In this Appendix we provide an alternate formulation to deriving the properties of the QHSL states discussed in Section.~\ref{section:abelian_Z2_QHSL}. 
We begin with the low energy effective theory in  the form 
 \be
 {\cal L} = {\cal L}(f, a+A) + {\cal L}(b_\alpha, -a) 
 \ee
 where $A$ is the external probe $U(1)$ gauge field while $a$ is the internal $U(1)$ gauge field. 
 
 For what follows it is useful to  generalize the above theory and consider a continuum Lagrangian in terms of these fields and allow placing it on an arbitrary space-time manifold with metric $g$. 
 Then to be precise we should view $A$ as a spin$_c$ connection, and not a $U(1)$ gauge field\footnote{A spin$_c$ connection is like a $U(1)$ gauge field but its field quantization is modified to 
$ \int \frac{da}{2\pi} = \int \frac{w_2[TM]}{2} $ on every oriented closed 2-cycle. $w_2[TM]$ is the second Stiefel-Whitney class of the manifold.  Viewing $A$ as a spin$_c$ connection enables defining fermions on arbitrary 3-manifolds without specifying the spin structure. See, eg, Ref. ~\onlinecite{senthil2019duality}}. The internal gauge field $a$ however is an ordinary $U(1)$ gauge field. Note that $a+A$ is then a spin$_c$ connection so that the fermionic parton $f$ couples to a spin$_c$ connection as it should. 

Now consider the state where $f$ fill their band with Chern number $C$. We can describe their low energy theory using $C$ $U(1)$ gauge fields $\alpha_I$ ($I = 1,..., C$): 
\be
{\cal L}_{eff}[f, a+A) = - \sum_I \frac{1}{4\pi} \alpha_I d\alpha_I + \frac{1}{2\pi} (a + A) \sum_I d\alpha_I 
\ee
In contrast to the main text where we integrated out $a$ here we will  integrate out $\alpha_I$. This will make some properties more transparent.  The result is 
\be
{\cal L}_{eff}[f, a+A) =  \frac{C}{4\pi} (a+A) d (a+A) + 2C (CS[g])
\ee
The last term is a gravitational Chern-Simons term which will contribute to the thermal Hall effect.  Combining this with the spinon part we get a useful effective action
\beq
{\cal L}_{eff} & = & {\cal L} (b_\alpha, a) + \frac{C}{4\pi} (a+A) d (a+A) + 2C (CS[g]) \\
& = & {\cal L} (b_\alpha, a) +  \frac{1}{4\pi}  ada + \frac{C}{2\pi} Ada  + C (CS[A,g])
\eeq
Here $CS[A,g] = \frac{1}{4\pi} AdA + 2CS[g]$ is a combined Chern-Simons term appropriate for a spin$_c$ connection.  This term alone contributes a background electrical Hall conductivity $\sigma_{xy} = C$, and a thermal Hall conductivity $\kappa_{xy} = C$. This should be added to the Hall conductivities of the remaining dynamical theory ; in the examples we are interested in there will in fact be no additional contributions from the dynamics, and so $\sigma_{xy} = \kappa_{xy} = C$.  

Let us specialize to states where the spinons are paired into a spin singlet state. This is the QHSL state and is described by the TQFT Lagrangian 
\be
{\cal L} =   \frac{C}{4\pi}  ada  - \frac{1}{\pi} \beta da + \frac{C}{2\pi} Ada  + C (CS[A,g])
\ee
where $\beta$ and $a$ are both ordinary $U(1)$ gauge fields.

The TQFT has a $K$-matrix
\be
\left(
\begin{array}{cc}
-C  &   2   \\
2  &      0  
\end{array}
\label{eq:K_matrix_append}
\right)
\ee
and a charge vector $q = (C, 0)$. Clearly there is four-fold ground state degeneracy on a torus, and there are 4 distinct quasiparticle sectors. Further clearly as mentioned above this $K$-matrix theory has no contribution to either $\sigma_{xy}$ or $\kappa_{xy}$. 

Define the quasiparticle $e$ with $l = (1, 0) $: this  has 
self-statistics $\theta_e = 0$, and an electric charge $Q_e = 0$. The other quasiparticle $m$ with $l = (0, 1)$ has statistics $\theta_m = \frac{\pi C}{4}$, and electric charge $Q_m = \frac{C}{2}$. 
Their mutual statistics $\theta_{em} = \pi$.  

What about the global $SU(2)$ spin of these quasiparticles? Strictly speaking we should introduce a background $SU(2)$ gauge field to keep track of this but it is easy to infer the result through simpler arguments. Note that $e$ couples to $a$ with charge-$1$ and therefore should be identified as a ``dressed" version of  the bosonic spinon. In particular it will inherit its $SU(2)$ spin and hence have $S = 1/2$. The $m$ couples to $\beta$ and has $S = 0$. 

Note that  in using the `standard' rules of $K$-matrix CS theories to infer these results, it is important that all the Chern-Simons gauge fields are ordinary $U(1)$ gauge fields as we chose above. If instead we have spin$_c$ connections then we must remember that there is also a `bare' fermionic  statistics of the corresponding quasiparticle that must be added to the ones coming from the Chen-Simons theory.

One may ask why we identify $e$ with $b_\alpha$ and not with $f$. The point is $f$ couples to $a+ A$ and not just to $a$. If we insist on working with $f$ then we should neutralize its electric charge by binding to a physical electron which is (of course) equivalent to working with the bosonic spinon. 

It is interesting to consider some special values of $C$. 

$C = 0$ is the usual $Z_2$ spin liquid. For $C = 1$, the  full theory is equivalent to $U(1)_4$ with a charge vector $2$, as is readily seen by the change of variables $a \rightarrow a - 2\beta$ followed by integrating out the $a$-field. 

For $C = 8$, we have $\theta_m = 0, ~mod~2\pi$, and $Q_m = 4$. Thus we can just regard the theory as a regular $Z_2$ gauge theory, together with a background $\sigma_{xy} = \kappa_{xy} = 8$.  We can just regard this background as a  bosonic $E_8$ state. 

For $C = 4$, $\theta_m = \pi$, and $Q_m = 2$. We can still regard this as a standard $Z_2$ gauge theory but shift notation $m \leftrightarrow \epsilon$. The quasiparticle with $l = (1, 1)$ is, in the new notation, the $m$ particle and has $Q_m = 2, S = 1/2$. We can remove the charge by binding a physical Cooper pair but not the spin. Thus we have an unusual $Z_2$ spin liquid where both $e$ and $m$ have spin-$1/2$.  Though unusual, this is not forbidden here. It is anomalous in a time reversal invariant system which our system is not, and so there is no problem.

\section{Details of the quantum Hall valley liquids}
\label{app:qhvl}
To better track the fact that the two valleys have the opposite Chern number, it is easier to use a rotating frame technique to reproduce the slave boson theory.   We use the followng parton theory\cite{sachdev2009fluctuating}:

\begin{equation}
	c_{i;a}=R_{i;a \alpha} \psi_{i;\alpha}
	\label{eq:rotation_frame_parton}
\end{equation}
where $a=+,-$ is the physical valley index.  $\alpha=1,2$ is a pseudo-valley index. $R_i$ is a $SU(2)$ matrix, which rotates the valley index.  The above parton theory has a $SU(2)$ gauge structure:

\begin{equation}
	R_i \rightarrow R_i U^\dagger_i,\ \ \Psi_i \rightarrow U_i \Psi
\end{equation}
where $\Psi=(\psi_+,\psi_-)^T$ and $U_i \in SU(2)$. 

In the flat band limit, the ground state is valley polarized. We can define an order parameter as $\vec n_i= C^\dagger_i  \vec{\sigma} C_i$, where $C=(C_+,C_-)^T$ and $\vec \sigma$ is Pauli matrix in the valley space.  We also define $\vec{m_i}=\Psi^\dagger_i \vec \tau \Psi_i$, where $\vec \tau$ denotes Pauli matrix in the pseudo-valley space. It is easy to show that:

\begin{equation}
	\vec{n}_i=   \mathcal{R}_i \vec m_i 
	\label{eq:rotate}
\end{equation}

The $\mathcal{R}_{i}$ is $3 \times 3~$ $SO(3)$ rotation matrix corresponding to the $2 \times 2~$ $SU(2)$ rotations:
\begin{equation}
\mathcal{R}_{i}^{\alpha \beta} = \frac{1}{2} \mbox{Tr} \left( R_{i}^\dagger \sigma^\alpha R_{i} \tau^\beta \right)
\end{equation}

Eq.~\ref{eq:rotate} means that the $\mathcal{R}_i$ rotates the order parameter in the valley space.  We always assume that  the mean field ansatz of $\psi_1$ has the same Chern number as $c_+$ and $\psi_2$ has the same Chern number as $c_-$.  Time reversal acts as:  $C \rightarrow \sigma_x C,  R \rightarrow \sigma_x R \tau_x,  \Psi \rightarrow \tau_x \Psi$.   The physical charge is carried by $\Psi$ and $R$ boson is neutral.

In the flat band limit, the valley polarized state can be reproduced by the following ansatz:  $\langle R_i \rangle=I$ and $\vec m_i=(0,0,\pm 1)$.  Basically the fermion $\Psi$ has the full pseudo-valley polarization along the direction of $\tau_z$.  In the following we always assume that $\vec m_i=(0,0,1)$ and the time reversal symmetry is broken.  This does not necessarily mean a valley polarization, which needs condensation of $R_i$.   A valley disordered phase can be constructed by requiring $\langle R_i \rangle=0$.  $R_i$ can be parametrized as

\begin{equation}
	R_i= \left(\begin{array}{cc}
	z_{i;+} &  -z_{i;-}^* \\
	z_{i;-} &  z_{i;+}^*
	\end{array}\right)
\end{equation}
with the constraint $|z_{i;+}|^2+|z_{i;-}|^2=1$.

Let us assume that $\Psi$ is pseudo-valley polarized to have only $\psi_1$ component. We can redefine $f_i=\psi_{i;1}$.  Then Eq.~\ref{eq:rotation_frame_parton} can be rewritten as

\begin{equation}
	c_{i;a}= z_{i;a} f_i
	\label{eq:CP1_parton}
\end{equation}
where $a=\pm$. 

The valley FM order parameter is now $\vec n_i=z^*_{i;a} \vec{\sigma}_{ab} z_{i;b}$, which is just the standard CP$^1$ representation.  Eq.~\ref{eq:CP1_parton} also reduces to the slave boson parton theory and the $SU(2)$ gauge structure is higgsed down to $U(1)$ by $\vec{m}=(0,0,1)$.  Then we can follow our discussion on QHSL  to melting the valley FM order by letting $z_i$ forms a paired condensate: $\langle \epsilon_{ab} z_{i;a}z_{j;b} \rangle \neq 0$.  Given that $f=\psi_1$ forms a Chern insulator with Chern number $C$, the property of this phase is the same as the corresponding QHSL:  it has a quantized Hall conductivity $\sigma_{xy}=C \frac{e^2}{h}$ and has topological order with four anyons.   The difference in  this quantum Hall valley liquid (QHVL) is that the time reversal is spontaneously broken without any quantized valley polarization.   Specifically the time reversal breaking will generically induce some valley polarization. However the amount of valley polarization will vary continuously throughout this phase. This should be contrasted with the QHSL where the valley is maximally polarized and hence quantized throughout the phase.  Note that in the quantum Hall valley liquid. the time reversal breaking is manifested in the mean field ansatz  $\psi=\psi_+$.

In the discussion of QHSL and QHVL, we assume valley polarization and spin polarization respectively.  It is also possible to imagine a Chern insulator without neither spin nor valley polarization.

\section{Anomalous Jain sequences from CFL\label{append:jain_sequence}}
In quantum Hall systems, fractional quantum Hall states can be generated from CFL by adding effective magnetic field to composite fermions. Here we generalize this procedure to the CFL proposed in Section. 

From Eq.~\ref{biqhe} and Eq.~\ref{fs_a}, variation of $a_0$ leads to
\begin{equation}
	-\delta n +C\frac{da}{2\pi}+C \frac{d A}{2\pi}=0
\end{equation}

Therefore, the spinon Fermi surfaces can feel an effective magnetic flux:
\begin{equation}
	-\frac{da}{2\pi}=\Phi-\frac{1}{C}\delta n
\end{equation}

At fix density, $\frac{da}{2\pi}=-\Phi$ and we will have quantum oscillations when applying external magnetic field.  When $n=1+\delta n=-\tilde C \frac{da}{2\pi}$, $f_\sigma$ can be in an IQHE states. Without considering spin polarization, we need $\tilde C=2p $ with $p \in Z$.   The condiction is $n=\tilde C\left(\Phi-\frac{1}{C}(n-1)\right)$, leading to $n=\frac{C\tilde C}{C+\tilde C} \Phi+\frac{\tilde C}{C+\tilde C}$.    We have a sequence of anomalous Landau fans starting from a QH state at $n=1$ with $\sigma_{xy}=\frac{C \tilde C}{C+\tilde C}$.  $\tilde C=\infty$ corresponds to $da=0$ and this is a CFL state  along the line with $n=C \Phi+1$.

The Hall conductivity can again be easily understood from Ioffe-Larkin rule $\rho_c=\rho_b+\rho_f$. We keep $b$ in the IQHE state with $\sigma_{xy}^b=C$, while letting $f$ in another IQHE states under the effective magnetic field $\sigma_{xy}^f=\tilde C$.  As a result, $\rho^c_{xy}=\frac{1}{C}+\frac{1}{\tilde C}$ and thus $\sigma^c_{xy}=\frac{C\tilde C}{C+\tilde C}$.   The case with $\tilde C=-C$ needs special treatment. In this case, $\rho^c_{xx}=\rho^c_{xy}=0$ because the Hall conductivities from $b$ and $f_\sigma$ cancel each other. This is a superconductor at strong magnetic field with $\Phi=\frac{1}{\tilde C}$!   In this paper we consider a spinful Chern band with fixed Chern number, then the sequences for both $\tilde C>0$ and $\tilde C<0$ are possible and they are generated by opposite magnetic field $B$. In moir\'e materials,  opposite Chern numbers will be selected by opposite sign of magnetic fields through opposite valley polarization. As a result, only one half of the Jain sequences discussed here are possible, which depends on the sign of valley Zeeman coupling.

Next we derive effective action for these states. The action for the IQHE states from $f_\sigma$ is
\begin{equation}
	L_{f,IQHE}=-\sum_{I=1}^{|\tilde C|}\frac{sign(\tilde C)}{4\pi}\alpha_I d \alpha_I-\sum_{I=1}^{|\tilde C|} \frac{1}{2\pi}a d \alpha_I
	\label{eq:f_IQHE_Jain}
\end{equation}

The final action is a sum of Eq.~\ref{biqhe} and Eq.~\ref{eq:f_IQHE_Jain}. Inegration of $a$ locks $\beta_1=\sum_{I=1}^{\tilde C}\alpha_I -\beta_2$.  It is easy to get

\begin{equation}
	L=\frac{C}{4\pi}\beta_2 d \beta_2-\sum_{I=1}^{|\tilde C|}\frac{sign(\tilde C)}{4\pi}\alpha_I d \alpha_I -\frac{1}{2\pi}\sum_{I=1}^{|\tilde C|} \beta d \alpha_I+\frac{1}{2\pi} \sum_{I=1}^{|\tilde C|}A d \alpha_I
\end{equation}

We have a $K$ matrix with dimension $|\tilde C|+1$. We find that $|Det K|=|C+\tilde C|$ and the chiral central charge is $\tilde C-1$.

When $\tilde C=-C$, $Det K=0$ and there must be a gapless mode.    For $C=2$, with redefinition $\alpha_c=\frac{1}{2}(\alpha_1+\alpha_2)$, $\alpha_s=\frac{1}{2}(\alpha_1-\alpha_2)$ and $\beta=\frac{1}{2}(\alpha_1+\alpha_2)-\beta_1$, the action can be rewritten as
\begin{equation}
	L=\frac{2}{4\pi}\beta d \beta+\frac{2}{4\pi}\alpha_s d \alpha_s +\frac{2}{2\pi} A d \alpha_c
\end{equation}

From $q_1\alpha_1+q_2\alpha_2+q_{\beta_1}\beta_q=q_c \alpha_c+q_s\alpha_s+q_\beta \beta$ we can get the charge transforms as $q_c=q_1+q_2+q_{\beta_1}$, $q_s=q_1-q_2+q_{\beta_1}$ and $q_\beta=-q_{\beta_1}$.  Here $\alpha_c$ is a gapless mode corresponding to the goldstone mode of the superconductor. Because the smallest $q_c$ is $1$, the fundamental flux for this superconductor is $h/2e$.   Actually the above action is the same as that for the $d+id$ superconductor.  For general $C$, the superconductor at $\tilde C=-C$ has a fundamental vortex with flux $\frac{h}{C e}$.

\section{Ground state degeneracy of the non-Abelian QHSL\label{append:counting_anyons}}

In this section we show that the non-Abelian QHSL proposed in Section.~\ref{section:CFL} has more anyons than a simple Ising TQFT or Pfaffian state, which has three anyons.

Let us consider $C=1$ for simplicity. The non-Abelian QHSL has a topological order ($U(1)_4\times TC \times $ Ising )$/Z_2$ (here TC refers to Toric Code/$Z_2$ gauge theory). The action for the Abelian TQFT part is

\begin{align}
	&L^c_{Abelian-TQFT}=-\frac{4}{4\pi} \alpha_1 d \alpha_1+\frac{2}{2\pi}A d \alpha_1+\frac{2}{2\pi}a d\alpha_2
	\label{eq:QHSL_abelian-TQFT}
\end{align}

 There are three gauge fields $\alpha_1, \alpha_2, a$ with charge $q_1,q_2, q_a$. As argued in Section.~\ref{section:CFL}, the Ising anyon is bound to the charge with $q_1+q_2=1$ mod 2.   Topological order $U(1)_4\times TC \times $ Ising has $4\times 4 \times 3=48$ anyons. Naively we may expect the number of anyon for ($U(1)_4\times TC  \times $ Ising )$/Z_2$ is $24$. However, in the following we will show that there is double counting and there are only $18$ anyons.

 Because the $m$ particle generated by $(q_1,q_2,q_a)=(0,0,1)$ does not couple with the Ising and the $U(1)_4$ sector, we can focus on the $q_a=0$ case first.  We will show that there are 3 anyons in the $q_2=0$ sector and 6 anyons in the $q_2=1$ sector when $q_a=0$.

When $q_2=0$, all of the anyons are from the bosonic Pfaffian state described by ($U(1)_4\times $Ising)$/Z_2$.  Each anyon is a bound state of one anyon in the Ising TQFT ($I,\psi, \sigma$) and an anyon in the $U(1)_4$ sector with charge $q_1$.  $\sigma$ bounds to odd $q_1$ and $I,\psi$ bound to even $q_1$.  Therefore, we have six anyons: $\sigma_1=(\sigma,q_1=1)$, $\tilde \sigma_1=(\sigma,q_1=3)$, $I_1=(I,q_1=0)$, $\tilde I_1=(I,q_1=2)$, $\psi_1=(\psi,q_1=0)$, $\tilde \psi_1=(\psi,q_1=2)$.   It is easy to show that the self statistics is $\theta_{\sigma_1}=\theta_{\tilde \sigma_1}=\frac{3}{8}\pi$, which is a sum of the self statistics of $\theta$ ($\theta_\sigma=\frac{1}{8}\pi$) and $q_1=1,3$ anyon in the $U(1)_4$ sector ($\theta_{q_1}=\frac{q_1^2}{4}\pi$). Meanwhile $\tilde I_2,\psi_1$ are fermions and $I_1,\tilde \psi_1$ are bosons.  We can also check that the mutual statistics between $\tilde \psi_1$ and the other anyons is $0$ mod $2\pi$, which means that both $I_1$ and $\tilde \psi_1$ are trivial anyons. Similarly we can check that $\tilde I_1$ and $\psi_1$ are equivalent. $\sigma_1$ and $\tilde \sigma_1$ are also indistinguishable.   As a result, there are only three anyons $I_1, \psi_1, \sigma_1$ in the $q_1=0$ sector (with the assumption that $q_a=0$).

Next we turn to the $q_2=1$ sector. Now $\sigma$ is bounded to even $q_1$ and $I,\psi$ are bounded to odd $q_1$.  Similarly we have six anyons: $\sigma_2=(\sigma,q_1=0)$, $\sigma_3=(\sigma,q_1=2)$, $I_2=(I,q_1=1)$, $I_3=(I,q_1=3)$, $\psi_2=(\psi,q_1=1)$, $\psi_3=(\psi,q_1=3)$. We find that $\theta_{\sigma_2}=\frac{1}{8}\pi$ and $\theta_{\sigma_3}=\frac{9}{8}\pi$, thus $\sigma_2$ and $\sigma_3$ are different.  Although the self statistics of  $I_2$ and $I_3$ are the same ($\theta_{I_2}=\theta_{I_3}=\frac{1}{4}\pi$), the fusion rules $I_2 \times I_2=\psi_1$ and $I_2\times I_3=I_1$ show that they are different. Similarly we can argue that $\psi_2$ and $\psi_3$ are different.

Combining the $q_2=0$ and $q_2=1$ sectors, there are in total $9$ anyons in the $q_a=0$ case.  Now we can choose to attach the $m$ particle generated by $(q_1,q_2,q_a)=(0,0,1)$. In total there are 18 anyons.

\end{document}